\documentclass[aps,prb,twocolumn]{revtex4-2}

\usepackage{amsmath}
\usepackage{graphicx}
\usepackage{dcolumn}
\usepackage{bm}
\usepackage[colorlinks=true, citecolor=blue, linkcolor=blue, urlcolor=blue]{hyperref}
\usepackage{natbib}
\usepackage{nameref}  

\begin{document}

\title{\textbf{Bismuth Films on EuO(111) as a Platform for Proximity-Induced Topological States}}

\author{Subham Naskar}
\author{Sujit Manna}%
 \email{Contact author: smanna@physics.iitd.ac.in}
\affiliation{Department of Physics, Indian Institute of Technology Delhi, New Delhi, India.}
  
\date{\today}

\begin{abstract}
Interfacing two-dimensional bismuth with a magnetic layer provides a promising route towards realizing higher-order topological phases. In particular, bismuthene on a ferromagnetic insulator substrate has been theoretically proposed by \citet{Chen2020} as a universal platform for magnetic second-order topological insulators. Here, we report the experimental realization of epitaxial bismuth films grown on the ferromagnetic insulator EuO(111). Using high-resolution scanning tunneling microscopy, we observe atomically ordered bi-layer bismuth with a (012)-oriented quasi-square lattice, corresponding to a stabilized $\alpha$-phase bismuthene. The resulting film is exceptionally flat compared to conventional metallic films, reflecting the intrinsic two-dimensional nature of the Bi(012) phase. Tunneling spectroscopy(STS) reveals a robust energy gap of about 400 meV in the local density of states, consistent with a quantum spin Hall insulating phase persisting up to room temperature. Spatially resolved STS further identifies enhanced edge-localised states at the island boundaries. Complementary low-temperature magnetotransport measurements on proximity-coupled ultrathin Bi films exhibit linear magnetoresistance and a Hall sign reversal, indicative of quantum-confinement-driven surface-dominated transport. Our results establish bismuthene–magnetic-insulator heterostructures as a viable experimental platform for realizing magnetically tunable topological phases, providing a critical step toward the observation of higher-order topology in two dimensions.
\end{abstract}

\maketitle

\clearpage
\section{Introduction}
\noindent Topological phases of matter in two dimensions have attracted sustained interest due to their potential for dissipationless transport and robust quantum functionality \cite{ren2016topological,chang2023colloquium,hsu2019topology,Hasan2010}. In quantum spin Hall (QSH) insulators, one-dimensional helical edge channels are protected by time-reversal symmetry, enabling resistance-free conduction immune to elastic backscattering \cite{Qi2011,kane2005quantum}. More recently, the concept of higher-order topology has generalised this paradigm, predicting lower-dimensional boundary modes such as corner states in two-dimensional systems \cite{Benalcazar2017,Benalcazar2017b,Song2017,Song2017,Schindler2018,Schindler2018b}. Incorporating magnetism into such topological platforms is particularly compelling, as it enables controlled symmetry breaking and access to novel phases including magnetic second-order topological insulators (SOTIs), which host localized zero-dimensional states with potential applications in spintronics and quantum information devices \cite{Liu2017,yang2025magnetic}.

\noindent Bismuthene, a two-dimensional monolayer of Bi atoms, is a promising material platform owing to its strong spin–orbit coupling, which promotes band inversion and large topological gaps \cite{YanezParreno2025,Murakami2006}. Previous experimental efforts to realize robust two-dimensional QSH phases in bismuth-based systems have largely focused on epitaxial Bi films and bilayers grown on nonmagnetic substrates, where signatures of topological edge states have been reported but remain highly sensitive to substrate hybridization and structural phase stability \cite{Reis2017,liu2011stable}. While Bi(111) bilayers and related allotropes exhibit nontrivial band topology driven by strong spin–orbit coupling, achieving a large, well-isolated bulk gap together with controlled symmetry breaking has proven challenging, limiting access to more exotic topological phases \cite{wada2011localized,murakami2008quantum}. In this context, recent theoretical advances, particularly by Chen et al., establish a universal route to magnetic second-order topology by introducing time-reversal symmetry breaking into an existing QSH insulator via magnetic proximity \cite{Chen2020}. Specifically, bismuthene on a ferromagnetic insulator such as EuO(111) is predicted to provide an ideal platform where strong SOC and sizable exchange coupling open a boundary gap (83 meV), while preserving bulk topology, leading to the emergence of protected corner states within the gapped edges. Importantly, this system offers a unique mechanism whereby the topological phase can be tuned across a transition from a first-order (QSH) to a second-order topological insulator through the onset of magnetic order, effectively controlled by the magnetic phase transition of the substrate \cite{Ezawa2018}. 
Complementarily, theoretical analysis of in-plane magnetization in bismuthene demonstrates that magnetic anisotropy plays a decisive role: the helical edge modes of the QSH phase become gapped under in-plane exchange fields, and a second-order topological phase emerges with localized corner states carrying fractional charges, whose spatial distribution is tunable by magnetization direction \cite{han2022plane}. Together, these works identify bismuthene–magnetic-insulator heterostructures as a uniquely versatile system in which the interplay of spin–orbit coupling \cite{liu1995electronic}, lattice symmetry \cite{schindler2018higher}, and magnetic proximity \cite{du2026strain} enables a controllable transition from first-order topology to higher-order phases. Despite this compelling theoretical framework, the experimental realization of epitaxial bismuthene on a magnetic insulating substrate and the direct microscopic validation of the predicted gapped edges and emergent boundary states have not yet been achieved.

Here we report the experimental realization of epitaxial bismuthene on EuO(111) and establish it as a viable platform for magnetic higher-order topology. Using high-resolution scanning tunneling microscopy and spectroscopy (STM/STS), we observe atomically-ordered bismuthene with a (012)-oriented quasi-square lattice corresponding to a rare $\alpha$-phase stabilized by the magnetic substrate. Spatially resolved tunneling spectroscopy reveals a clear energy gap in the local density of states, consistent with a QSH insulating phase, along with distinct edge-localized states at the boundaries of bismuthene islands. While direct observation of corner states is constrained by measurement temperatures above the EuO ferromagnetic ordering temperature (60 K), our results demonstrate the essential ingredients required for a magnetic SOTI phase. Complementary low-temperature magnetotransport measurements on ultrathin Bi films proximitized by EuO show linear magnetoresistance and Hall sign reversal, indicating quantum confinement induced surface dominated transport. Together, our findings bridge the gap between theoretical proposals and experimental realization, and provide a robust platform for future exploration of magnetic higher-order topological phases.

\section{Experimental Methods}

\noindent Epitaxial EuO(111) thin films of controlled thickness were grown on highly conducting Si(100) substrates in an ultrahigh vacuum (UHV) chamber (base pressure $\sim 4 \times 10^{-7}$ mbar) using pulsed laser deposition (PLD), following our earlier protocols \cite{sinha2025magnetic}. Before deposition, Si(100) substrates were cleaned in situ by repeated annealing cycles to obtain atomically clean surfaces. A KrF excimer laser ($\lambda = 248~\mathrm{nm}$, repetition rate = 5 Hz, pulse duration = 20 ns) was used to ablate a stoichiometric EuO target (99.999\%) with an energy density of $\sim 400$ mJ per pulse. The film thickness was controlled by the number of laser pulses. During growth, an oxygen flow rate of 1.5 sccm was maintained, corresponding to a chamber pressure of $1.1 \times 10^{-4}$ mbar. Structural and magnetic optimization of EuO(111) films was carried out using X-ray diffraction (XRD), X-ray reflectivity (XRR), and DC magnetization measurements. To prevent surface oxidation of UHV-grown films, special precautions were taken during transfer to the other chamber and subsequent processing.

\noindent Bismuth thin films were deposited in situ on EuO/Si(100) using a high-purity Bi target (99.999\%) in a UHV deposition system (base pressure $\sim 2 \times 10^{-8}$ mbar). Prior to the Bi deposition, the EuO films were outgassed to remove any contamination. Post-deposition annealing was carried out at $150^\circ$C for 30 min to improve crystallinity. Film thickness and deposition rate (0.5 ML/min) was calibrated using combined XRR, atomic force microscopy (AFM), and scanning tunneling microscopy (STM) surface coverage analysis. Here, we define one bilayer (BL) as the areal density of Bi atoms corresponding to a pseudocubic Bi(012) plane ($2.8 \times 10^{22}$ atoms.$\mathrm{cm^{-3}}$). For indexing the X-ray diffraction (XRD) peaks of Bi, we adopt the lattice convention reported in \cite{nagao2004nanofilm}, with bulk lattice constants a = b = 4.54 Å, and c = 11.8 Å given in fig. S1(a). Structural characterization was performed using a PANalytical X-ray diffractometer with Cu–K$\alpha$ radiation ($\lambda = 1.540$ Å). Surface morphology and roughness were examined by AFM (Asylum Research MFP-3D) in tapping mode using AC240TS-R3 cantilevers. Raman spectroscopy measurements were conducted at room temperature using a Renishaw inVia Reflex system (532 nm excitation, $\sim 6$ mW, spot size $\sim 1~\mu$m$^2$). Chemical composition and oxidation states were analyzed by X-ray photoelectron spectroscopy (XPS) using a Kratos AXIS Supra system with a monochromatic Al–K$\alpha$ source (1486.6 eV).

\noindent Atomic-scale characterization was performed using room-temperature STM/STS in a UHV system equipped with active vibration isolation \cite{sinha2026imaging}. Before measurements, samples were outgassed insitu for 2 h and further annealed at $150^\circ$C for 90 min to remove surface contamination. STM tips were fabricated from polycrystalline W wire by chemical etching followed by in situ annealing. Differential conductance ($dI/dV$) spectra were acquired using standard lock-in techniques. DC magnetization and magnetotransport measurements were carried out using Quantum design MPMS3 and PPMS Evercool systems over the temperature range 2–300 K, with out-of-plane magnetic fields up to $\pm 7$ T.

\section{Results and Discussion}

\noindent Recent first-principles calculations predicted \cite{Chen2020} that bismuthene interfaced with EuO(111) hosts magnetic SOTI with symmetry-protected, spin-polarized corner states, representing one of the first realistic two-dimensional platforms for such phases. Unlike previously proposed systems such as graphdiyne, which are effectively spinless \cite{sheng2019two}. The Bi/EuO heterostructure leverages strong spin–orbit coupling in bismuth combined with exchange proximity from the ferromagnetic insulator to generate a sizable boundary gap and robust topological corner modes. This concept critically relies on the experimental ability to grow bismuthene and integrate with a magnetic insulator with well-defined long-range order, such that proximity-induced exchange interactions can drive the transition from first-order to higher-order topology. Motivated by this proposal, we first focus on the synthesis of high-quality EuO thin films with (111) termination, which provide in-plane magnetization, exhibiting a large exchange field and a suitable platform for realizing the predicted magnetic SOTI phase. As compared to other magnetic insulators like EuS, EuO has the virtues of a higher magnetic ordering temperature (T$_C$ = 70 K) and a large spin-splitting state of 0.6 eV \cite{santos2008determining}. We grown high-quality epitaxial EuO(111) film on Si(100), which serve as a template for stabilizing ultra-thin bismuth. X-ray diffraction measurements in fig. 1(b) show a single dominant EuO(111) peak with no secondary phases, confirming phase-pure, highly textured growth. The extracted lattice parameter, a = 5.14 Å is slightly expanded relative to the bulk value \cite{zhang2018high}, indicative of thin-film strain. Careful control of oxygen partial pressure suppresses the formation of parasitic $\mathrm{Eu_2O_3}$. The EuO(111) surface exhibits a hexagonal lattice with an in-plane Eu–Eu spacing of 3.594 Å shown in fig. 1(a), providing both crystalline registry and in-plane magnetization essential for proximity-induced exchange coupling.

\begin{figure*}[t]
  \includegraphics[width=\textwidth]{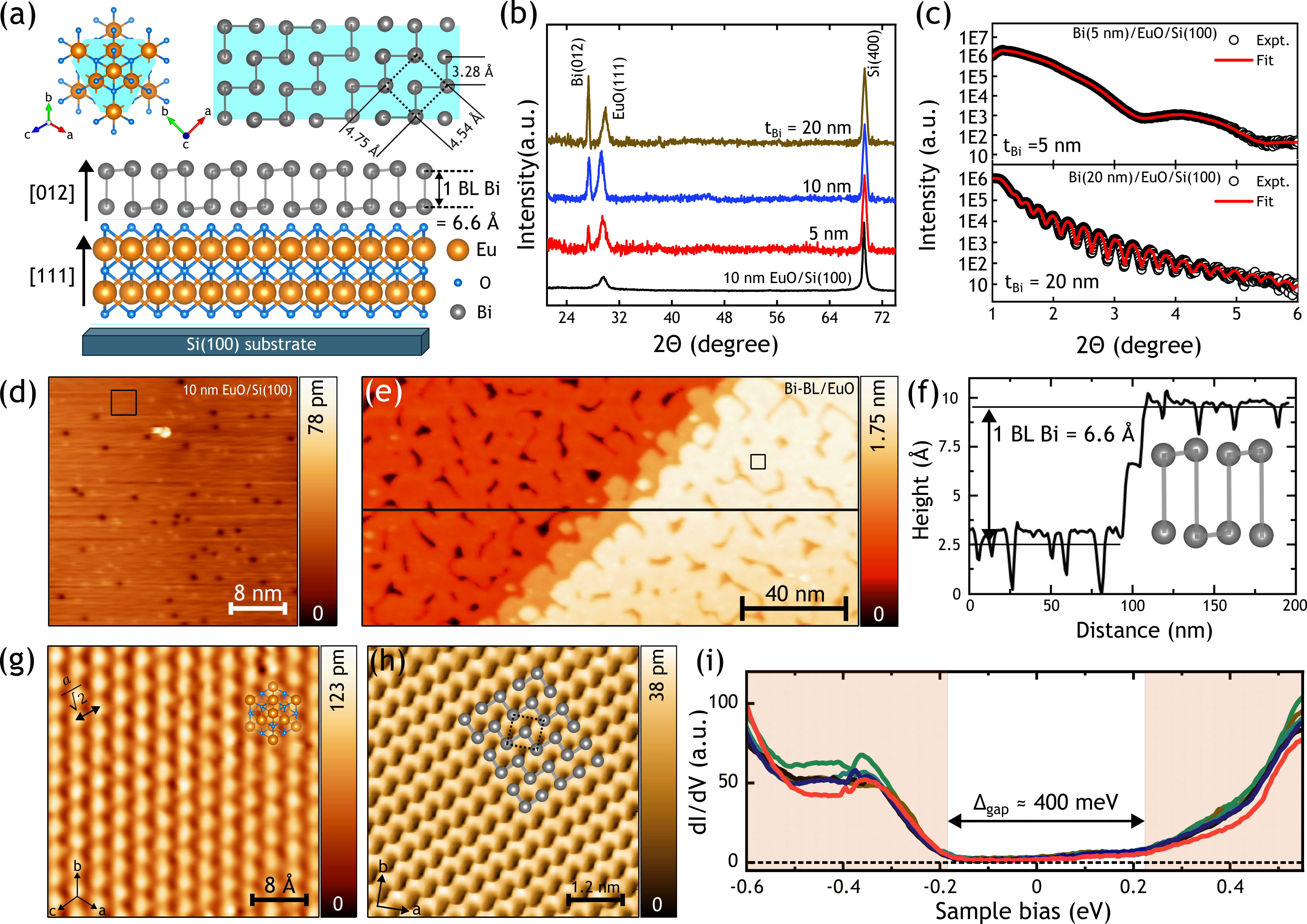}
  \caption{Structural characterization of 2D Bi/EuO heterostructures.
(a) Schematic illustration of the top views of EuO(111), exhibiting a plane shares hexagonal lattice. Bi(012) displays an in-plane quasi-square lattice with lattice constants of 4.75 $\mathrm{\AA}$ and 4.54 $\mathrm{\AA}$. A side view depicts bilayer Bi(012) stacked with EuO(111). 1 bilayer (BL) Bi along (012) corresponds to the thickness of 6.6~\text{\AA}. (b) Thickness-dependent XRD plot for Bi/EuO heterostructure with varying Bi thicknesses (t = 5 nm, 10 nm, 20 nm). (c) X-ray reflectivity (XRR) measurement and fitted curve for 5 nm and 20 nm Bi film. (d) STM topography image ($40 \times 40\,\text{nm}^2$) of 10 nm EuO(111) film surface (I$_t$ = 0.5 nA , V = 200 mV), showing flat terrace with oxygen vacancies as point defects. (e) Large-scale($200 \times 200\,\text{nm}^2$) STM topography image of bilayer bismuth film on EuO(111) with $\sim$ 95\% surface coverage ( I$_t$ = 50 pA, V = 0.8 V). (f) Line profile along the black line in (e), indicating the bilayer step height of Bi. (g) Atomically resolved STM topography image ($4 \times 4\,\text{nm}^2$) of EuO(111) film surface (I$_t$ = 1.0 nA, V = 150 mV) with hexagonal symmetry, the corresponding ball model is shown (top right). The zoomed-in region is marked by the black square in (d). (h) Atomically-resolved STM image of $6\,\mathrm{nm} \times 6\,\mathrm{nm}$ area on the quasi-square facet of Bi(012) film. The corresponding atomic model is overlaid. The imaged region is indicated in (e). ( I$_t$ = 0.5 nA, V = 100 mV). The scanned area is the black square box in (e). (i) Position-dependent differential conductance(dI/dV) taken on BL Bi film exhibits an energy gap of $\sim$ 400 meV.}
  \label{fig:structural properties}
\end{figure*}

\noindent Subsequently, we have optimized the controlled epitaxial growth of single-crystalline bismuth films of different thickness(t) grown on 10nm EuO(111). Unlike bulk Bi, which preferentially cleaves along the (111) plane, Bismuth (Bi) films on EuO prefer to grow in the unusual (012) orientation as estimated from thickness dependent XRD study (shown in Fig. 1b). Using UHV deposition, we achieve single-crystalline Bi films on EuO(111) with thicknesses ranging from single BL to several nanometers, enabling access to both ultra-thin and quasi-bulk regimes. The crystallographic orientation and thickness of Bi films are confirmed by combined XRD and X-ray reflectivity(XRR) measurements. Figure 1(b) shows that Bi grows preferentially along the (012) direction on EuO(111), while the underline EuO(111) peak remains dominant, indicating good crystallinity and interface quality. XRR study as depicted in fig. 1(c) provides precise thickness calibration for films in the 5–20 nm range. 
Fig. 1(a) shows the stacking orientation of EuO (111) and Bi (012), where thickness of single bilayer (BL) bismuth is 6.6 Å. Bismuth thin film deposition is complicated task due to its low melting point (271.3\textdegree C), which leads to higher crystallization rates and low vapor pressure. Previous studies suggest that low deposition energy leads to (111)-oriented growth, however higher deposition energy leads to (012)-oriented growth \cite{Rodil2017}. Stanley et al. \cite{Stanley2012,Stanley2015} have reported that bismuth films exhibit a dominant [012] texture over [003] at 125°C, followed by the emergence of a bimodal grain size distribution. (012) phase of Bi forms quasi-square-shaped surface unit cell of lattice parameters of 4.75~\text{\AA} and 4.54~\text{\AA}, respectively. Here, Bi$\{012\}$ phase has a striking analogy to the puckered-layer structure of black phosphorus \cite{Takahashi1986}, except (i) the difference in the stacking sequence of these puckered layers and (ii) strong buckling in each (012) plane.

\noindent Raman spectroscopy (see Supplementary fig.~S2) further confirms the structural quality and phase purity of the Bi films. Bi films grown at different temperatures exhibit characteristic $E_g$ (in-plane) and $A_{1g}$ (out-of-plane) phonon modes. Films grown at higher temperature show reduced FWHM and a slight blueshift, indicating improved crystallinity, reduced disorder. Importantly, no signatures of oxide-related phases are detected, confirming the chemical integrity of the films.

\begin{figure*}[t]
  \includegraphics[width=\textwidth]{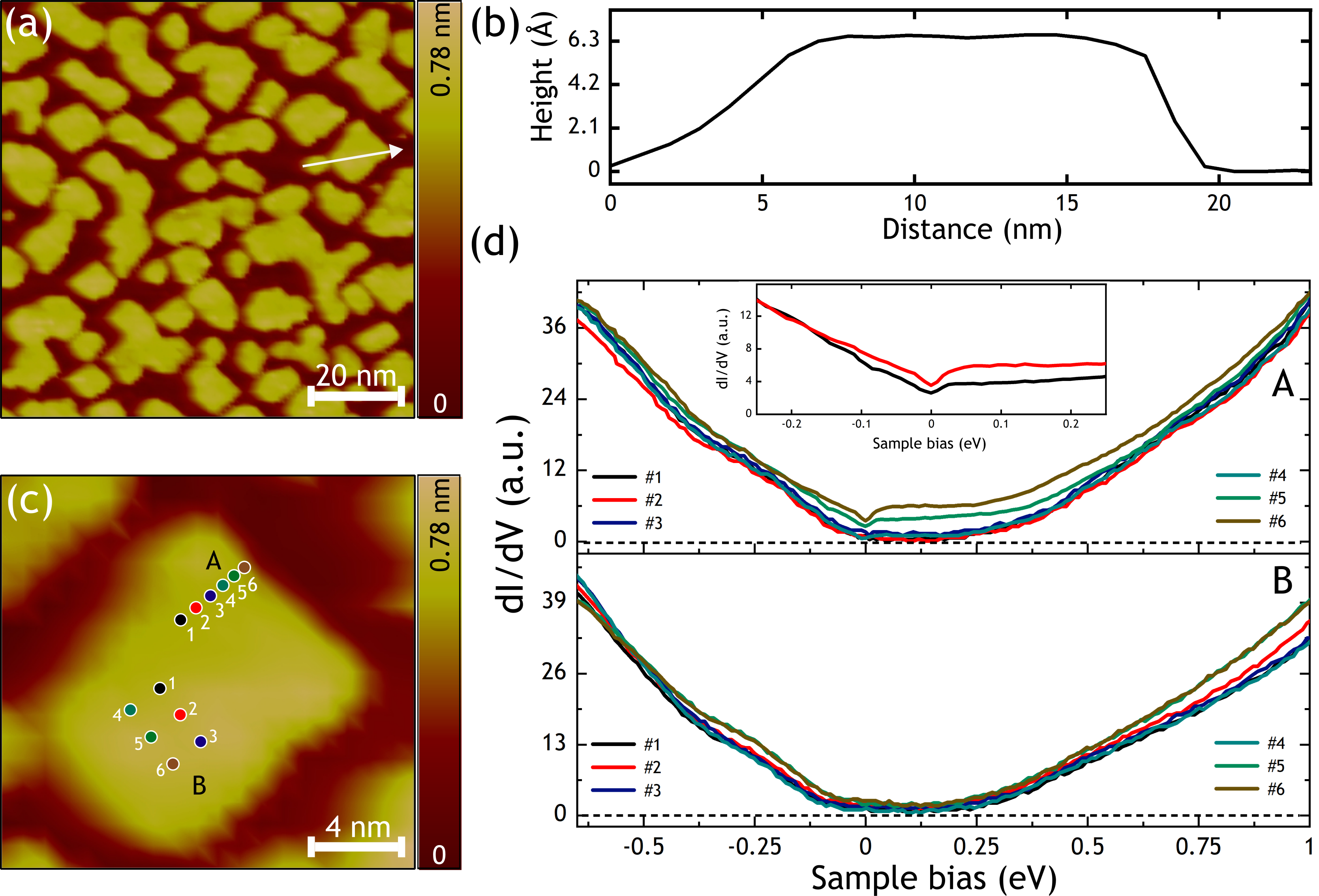}
  \caption{(a) Large-scale STM topography image ($100 \times 100\,\text{nm}^2$) of bilayer bismuth island grown on EuO(111) with area coverage of 60\% (I$_t$ = 100 pA, V = 0.65 V). (b) Height profile of a single bilayer bismuth island. The line scan is taken along the white arrow in (a), confirming a single bilayer Bi(012) step height. (c) Zoomed high resolution STM topography image ($20 \times 20\,\text{nm}^2$) of an isolated BL-island. with coloured markers indicating positions for spatially resolved spectroscopy. Two representative regions are highlighted: edge (A) and interior (B).
(d) Differential conductance (dI/dV), spectra acquired at positions A and B. A gap of $\sim$400 meV is observed in the island interior, while spectra near the edge show enhanced LDOS. Inset: Zoomed-in spectra at two adjacent edge positions reveal a zero-bias anomaly and increased density of states.}
  \label{fig:STM}
\end{figure*}

\begin{figure*}[t]
\includegraphics[width=\textwidth]{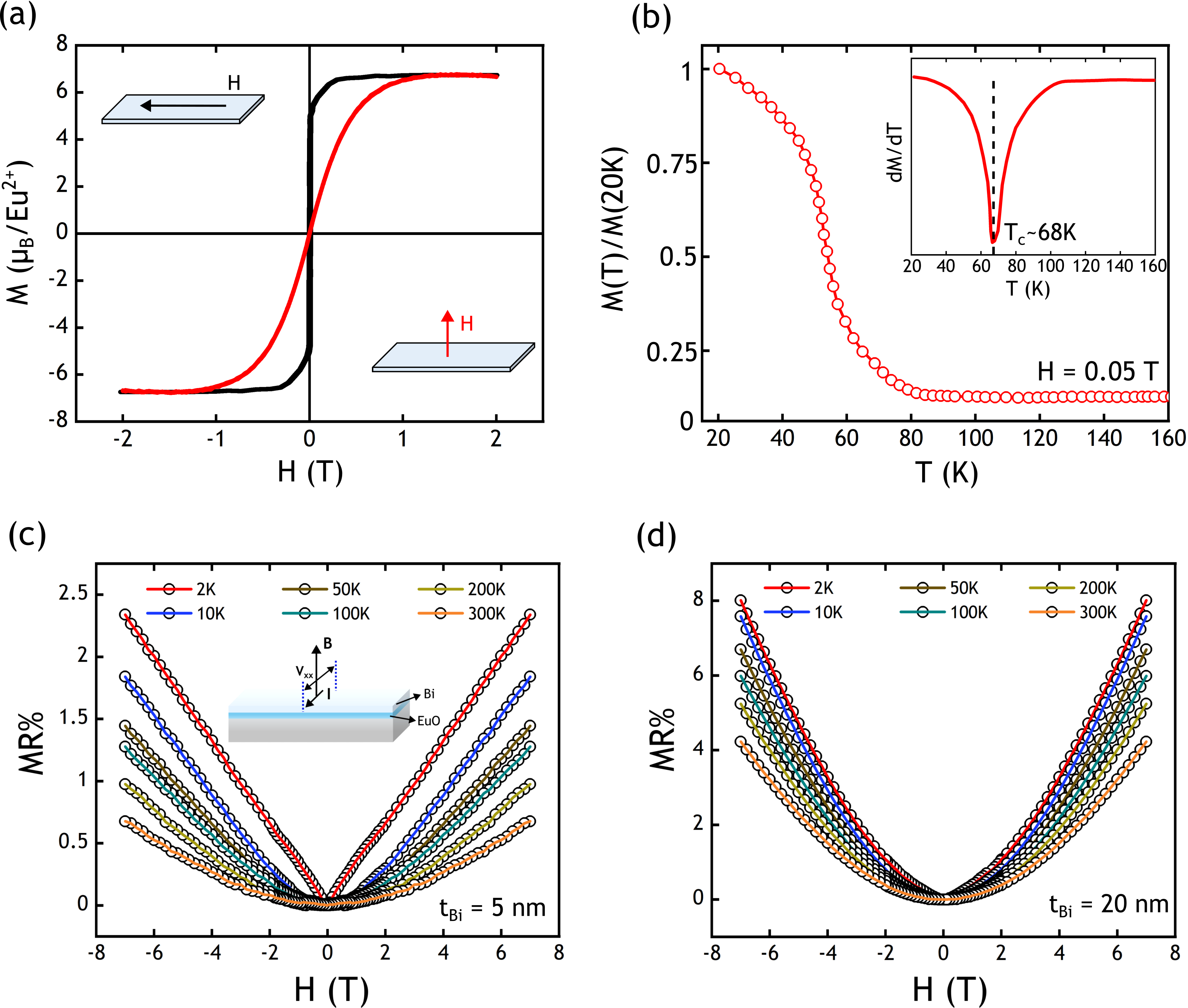}
\caption{Magnetization and Magnetotransport studies of 2D Bi/EuO heterostructure.
  (a) Magnetic-field-dependent magnetization measured at $T = 10~\mathrm{K}$ 
  with in-plane (black) and out-of-plane (red) magnetic fields respectively.
  (b) Magnetic anisotropy of the EuO underlayer as substrate, T-dependent magnetization of the Bi/EuO (111) heterostructure measured in field-cooling mode with $H = 0.05~\mathrm{T}$.
  Magnetoresistance versus applied magnetic fields (out-of-plane direction) at selected temperatures from 2 K to 300 K for Bi/EuO heterostructure of bismuth film thicknesses (c) 5 nm (d) 20 nm.
  \textit{Inset:} $dM/dT$ vs.\ $T$ plot showing $T_{\mathrm{C}}$ in fig. 3(b). Schematic representation of magnetoresistance measurement geometry in fig. 3(c).}
  \label{fig:magnetization and magnetotransport}
\end{figure*}

\noindent To understand the mode of growth for both EuO(111) and Bi layer on top, we investigate the surface structure of magnetic heterostructure using Scanning Tunnelling Microscopic (STM) technique. Real-space STM imaging reveals a predominantly flat surface with occasional dark features attributed to oxygen vacancies. Fig. 1(d) shows STM topograph of 10nm EuO film surface. Flat closed packed EuO(111) surface contains point defect appeared as black depressions, correspond to oxygen vacancies of Eu-rich films \cite{klinkhammer2014spin}. A critical concentration of oxygen defects are well-known in epitaxial EuO film that determine the metal insulator transition \cite{sinjukow2003metal}. As the samples are synthesized in an Eu-rich ambiance, the formation of oxygen vacancies is highly probable. Atomic-resolution images acquired on the area (without defect) denoted by small square box in (d) to identify atomic structure. Fig. 1(g) shows atomic-resolved STM image confirms well-defined hexagonal arrangement of Eu surface atoms at (111) surface. STM topograph reveals interatomic spacing 3.8 $\mathrm{\AA}$, equals to a/$\sqrt{2}$.

With the EuO(111) surface providing a structurally and magnetically well-defined template, bismuth growth proceeds via a bilayer-by-bilayer(BL) mode that yields large-area, uniform terraces. Fig. 1(e) shows large-area topographic image with wide terrace of bilayer Bi growth on EuO(111)/Si(100). The surface is uniformly covered with flat terrace with an approximate surface coverage of 95\%. The STM height profiles in fig. 1(f) shows a step height of ~6.6 $\mathrm{\AA}$, which matches that of the Bi(012) bilayer, confirming the structural unit along the trigonal [001] c axis. High-resolution STM imaging directly confirms the formation of a quasi-square lattice consistent with the Bi(012) surface, establishing the realization of bismuthene on EuO(111). The zoomed-in topograph in fig. 1(h) reveals a well-ordered atomic arrangement that agrees quantitatively with the expected unit cell \cite{nagao2004nanofilm,koch2024morphology}, as highlighted by the overlaid ball model. This quasi-square symmetry is a distinct signature of the Bi(012) orientation and clearly distinguishes it from the conventional Bi(111) structure. Notably, this structural phase corresponds to the $\alpha$-phase of bismuthene previously identified in epitaxial Bi films \cite{takahashi2022thickness,zhang2016semiconducting}, thereby confirming that our growth stabilizes this nontrivial allotrope. The ability to realize $\alpha$-phase bismuthene on a magnetic insulating substrate represents a key advance toward engineering proximity-driven topological phases in Bi-based heterostructures.

\begin{figure*}[t]
  \includegraphics[width=\textwidth]{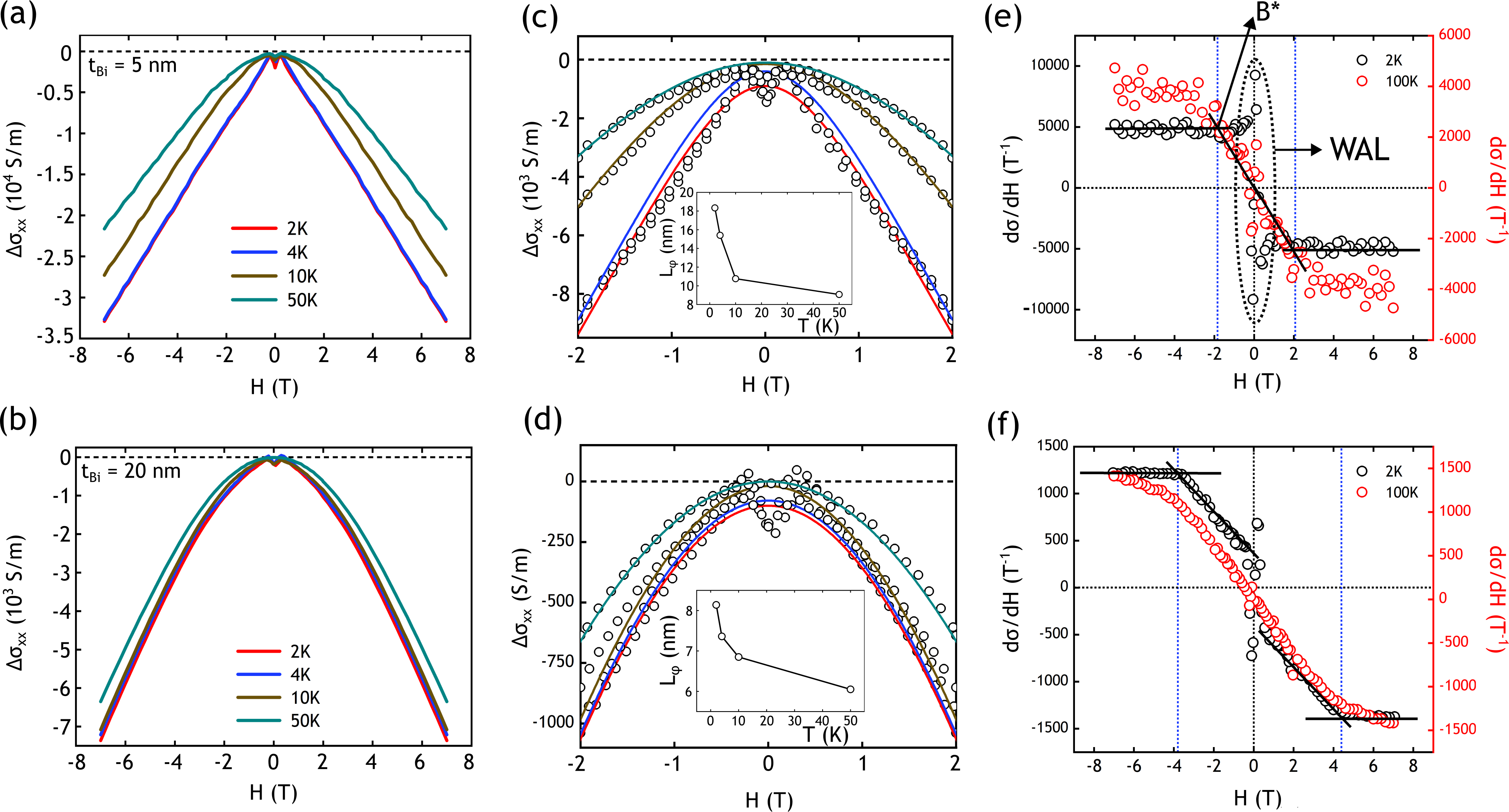}
  \caption{Weak antilocalization (WAL) effect and 2D magnetoconductance in Bi/EuO heterostructure.
  Magnetic field dependence of $\Delta \sigma_{xx}$ of Bi/EuO heterostructure for Bi thin film of thicknesses \textbf{(a)}
  5 nm \textbf{(b)} 20 nm. Zoomed-in view of (a) and (b) with HLN fitting plotted within the range of -2T to 2T for Bi film thickness \textbf{(c)}
  5 nm \textbf{(d)} 20 nm. Inset of both figures show temperature dependence of phase coherence length ($L_\phi$).
  \textbf{(e, f)} Variation of magnetic field derivative of conductivity with the applied magnetic field for $B_{\perp}$ to film plane (for temperatures at T = 2 K and 100 K). The critical field ($B^{*}$) is determined from the intersection of a straight line and horizontal line as depicted in fig. 4(e, f).}
  \label{fig:2D MC}
\end{figure*}

\noindent High resolution local density of states(dI/dV) measured in STS mode provide direct evidence of a large and robust energy gap in Bi(012) bismuthene on EuO(111), establishing its quantum spin Hall insulating character. The differential conductance spectra in fig. 1(i) consistently reveal a gap of ~400 meV that remains remarkably uniform across different spatial locations on the Bi bilayer, indicating high electronic homogeneity and minimal disorder. Notably, this gap persists at room temperature, underscoring its robustness and practical relevance. The magnitude and stability of the observed gap are in excellent agreement with prior reports of bismuthene on SiC by Reis et al. \cite{reis2017bismuthene}, widely recognized as a high-temperature QSHI platform. Our results thus demonstrate that bismuthene grown on a magnetic insulator retains a large topological gap while simultaneously enabling proximity-induced magnetism, a crucial requirement for realizing magnetic higher-order topological phases.

To access boundary and corner physics relevant to higher-order topology, we intentionally deposit submonolayer Bi on EuO(111) under controlled slow growth conditions, resulting in well-isolated bilayer islands with atomically sharp edges. Figure 2(a) shows a large-area STM image of such islands, covering about 60\% of the EuO(111) surface, indicating controlled nucleation and growth. The corresponding height profile in fig. 2(b) confirms a uniform step height of 6.6 $\mathrm{\AA}$, consistent with a single Bi(012) bilayer. A zoomed-in STM image in fig. 2(c)  identifies spatially resolved spectroscopy points extending from the island edge (region A) to the interior (region B). Differential conductance spectra in fig. 2(d) reveal a clear evolution of the local density of states, while spectra acquired in the interior exhibit a well-defined gap of ~390-400 meV, consistent with a QSH insulating phase in complete BL Bismuth, the gap progressively closes as the measurement position approaches the island edge, accompanied by a pronounced enhancement of in-gap states. This spatially resolved gap closing provides a signature for confined metallic channels localized at the boundaries of bismuthene, consistent with the emergence of 1D edge states expected for QSH systems \cite{schindler2018higher}. Owing to thermal drift at room temperature, full spatial mapping of the local density of states across the entire island was not feasible, which could have provided further confirmation. Nevertheless, the distinct spectroscopic contrast between edge and interior regions is consistently reproduced across multiple islands, yielding qualitatively similar behaviour. The observed edge-enhanced conductance and zero-bias features are qualitatively consistent with interacting helical channels exhibiting Tomonaga–Luttinger liquid behavior, as reported in bismuthene-based systems, thereby establishing the Bi/EuO heterostructure as a viable platform for exploring boundary and higher-order topological states. However, to unambiguously identify the higher-order topological phase and associated corner states, measurements at temperatures below the EuO Curie temperature (T $<$ T$_c$) are essential.

\noindent To establish the magnetic ground state of the heterostructure, we probe the magnetism and magnetotransport response of Bi(t)/EuO bilayers for t = 5 nm and 20 nm over a wide temperature range (2–300 K). Noting that direct detection of proximity-induced magnetism in bismuthene remains nontrivial and beyond the scope of the present measurements. DC magnetic characterization reveals that the Bi/EuO heterostructure retains robust ferromagnetism with strong in-plane anisotropy, as evidenced by hysteresis loops measured at 10 K shown in fig. 3(a), where the easy axis lies within the film plane with a coercive field (H\textsubscript{c}) of ~140 Oe, remanence (M\textsubscript{r}) of $5~\mu_{\mathrm{B}}$ per Eu$^{2+}$ atom and saturation magnetization (M\textsubscript{sat}) of ~$6.73~\mu_{\mathrm{B}}$ per Eu$^{2+}$ atom. In contrast, out-of-plane measurements remain unsaturated up to 1 T, indicating pronounced shape anisotropy. Temperature-dependent magnetization in fig. 3(b) shows a clear ferromagnetic transition at T$_C$ $\sim$ 68 K, consistent with bulk EuO, confirming that the magnetic substrate remains intact upon Bi deposition. This is a crucial prerequisite for realizing proximity-induced exchange coupling in bismuthene, as theoretical studies predict that below T $<$ T$_C$ , a moderate exchange field can gap out the helical edge states while preserving bulk band inversion, thereby driving the system into a magnetic second-order topological phase with a sizable boundary gap. The requirement that the exchange splitting remains smaller than the bulk gap ensures that the topological character is preserved, while enabling the emergence of spin-polarized corner states. Thus, the observed magnetic ground state in Bi/EuO provides the essential condition for tuning between 1D helical edge states and 0D corner modes, offering a controllable pathway to experimentally access higher-order topology via low-temperature spectroscopic probes. For example, topologically protected corner states can be probed by STS technique, which maps local density of states at liquid helium temperature. Unlike graphdiyne, Bi/EuO possesses spin-polarized corner states, which can be identified by using a ferromagnetic tip in STS. By tuning EuO substrate magnetism ($T < T_{\mathrm{C}}$), one can detect the transition between 0D corner states and 1D helical states.

\noindent Building on the established magnetic ground state, magnetotransport measurements provide a complementary probe of the electronic response of Bi films proximitized by the ferromagnetic EuO. We first examine the longitudinal resistivity  (see fig. S5a) using standard four-probe dc measurements on millimeter-scale devices, which establishes the baseline transport behavior across 2–300 K. The field-dependent magnetoresistance (MR), shown in figures 3(c) and 3(d) reveals a pronounced thickness dependency. In the ultrathin (t = 5nm) Bi(012) films, the MR is is strongly suppressed \(\sim2.4\%\) at 2 K and 7 T), which may attributed to enhanced surface and grain boundary scattering, and reducing the carrier mobility at lower thickness. Another aspect is quantum confinement effect at reduced thickness, further perturb electron-hole compensation, leading to diminished MR. In contrast, thicker films (20 nm) exhibit a significantly enhanced MR response, reaching values as high as \(\sim8\%\) at 2 K and 7 T. This enhancement in MR likely to be arises from the improved bulk-like electronic states with increasing thickness, resulting in higher carrier mobility and improved electron–hole compensation. Notably, the low-field MR cusp observed in thinner films (fig. 4a), together with a crossover from quadratic to linear field dependence, points to quantum transport dominated by confined states. The positive magnetoresistance with a sharp low-field cusp in a quantum spin Hall insulator reflects the breakdown of time-reversal symmetry–protected helical edge transport, serving as a key signature of its topological nature. However, it is absence in thicker films indicates a transition toward semiclassical bulk behavior. Kohler scaling (see fig. S5 b-c) further supports the coexistence of multiple carrier types, underscoring the complex interplay \cite{Marcano2010} between bulk and surface channels in the Bi/EuO heterostructure. 

\noindent To further elucidate the role of surface transport phenomenon, we analyze the low-field magnetoconductance (Fig. 4c-d) within the Hikami–Larkin–Nagaoka framework(HLN) formalism \cite{Hikami1980,Liu2012}, which captures the weak antilocalization (WAL) signature characteristic of systems with strong spin–orbit interaction. To evaluate WAL, we plot magnetoconductance (MC) data between -2 T and 2 T, in which magnetic field is low enough to let the electronic system stay in the WAL regime. The magnetoconductivity curve for 5 nm film at 2 K has a clear WAL cusp, which confirms the less elastic scattering and spin-orbit scattering in low magnetic field region. Whereas, in figures 4(c) and 4(d), the fitted curve agrees well with the experimental data. From HLN fitting, we calculate phase coherence length ($L_{\phi}$) at each temperature by the relation, $B_{\phi} = \frac{\hbar}{4 e L_{\phi}^{2}}$ \cite{He2011,Liu2011,Lu2011}.

\noindent The temperature dependence of $L_{\phi}$ is shown in inset of figs. 4(c) and 4(d). The phase coherence length decreases with increasing temperature, and coherent with the observation in magnetoconductance plot, which depicts WAL dip shrinks as temperature increases. At reduced thickness, $L_{\phi}$ is limited by 2D electron–electron (Nyquist) scattering. In dephasing mechanism, exceeding a certain limit of temperature causes phonon scattering more dominant than electron-electron interaction. The obtained value of the pre-factor ($\alpha$) remains within -0.2 to -1.2 for both the thicknesses. The increasing value of $\alpha$ with increasing film thickness describes that the contribution of topological surface states to the conduction is weak for thicker film. The transition from semi-classical to quantum is more pronounced in the field derivative of MC depicted in Fig. 4(e, f). Initially, $\frac{d\sigma}{dB}$ is proportional to B, which is indicative of $B^2$-dependent MC. As we increase B, $\frac{d\sigma}{dB}$ crosses critical field ($B^*$), and become independent of B. This implies that MC (or MR) follows linear-field dependence at a high magnetic field and quadratic field dependence at low magnetic field. This crossover vanishes at T = 100 K. All MC curves become quadratic in the entire applied magnetic field regime.
 
\noindent Further, we performed Hall measurements to determine the carrier type and carrier concentration in the field range of ±7 T at different temperatures. Figures S6(a) and (b) show the Hall resistivity from room temperature down to 2 K for the Bi films with thicknesses t = 5 nm and 20 nm as representative of the present study. At low field, $\rho_{xy}$ exhibits a linear field dependency whereas with higher field value, Hall resistivity slightly deviates from linear regime. The sign change in $\rho_{xy}$ from 20 nm to ultrathin limit of 7 BL for bismuth film is remarkable as one can expect that as the film thickness decreases, the surface-to-volume ratio increases, making surface carriers increasingly significant and potentially dominant. Furthermore, in ultrathin films, quantum confinement and finite-size effects can modify the positions of the bulk electron and hole bands, further elevating the role of the surface band. For the quantitative analysis of the density and mobility of charge carriers, we followed the well-known two-band model for fitting by considering two types of charge carriers (electron and hole) co-existence as the slope of $\rho_{xy}$ changes sign with film thickness in the field-range of 0–7 T. For $\rho_{xy}$ fitting, we employed two-band model fit described in supplementary material with fig. S6.

\noindent The carrier concentrations for both electrons and holes ($n_e$ and $n_h$), extracted from the fitting curve at 2 K are $n_e$ = $1.745 \times 10^{19}~\mathrm{cm}^{-3}$, and $n_h$ = $5.21 \times 10^{18}~\mathrm{cm}^{-3}$ for t = 5 nm. However, with increased thickness (t = 20 nm), at 2 K, $n_e$ = $1.34 \times 10^{18}~\mathrm{cm}^{-3}$, and $n_h$ = $6.7 \times 10^{17}~\mathrm{cm}^{-3}$. At ultrathin limit, carrier concentration increases as surface conduction becomes prominent. This value of carrier concentration is comparable with other Dirac semimetals such as Cd$_3$As$_2$ ($10^{16}$--$10^{17}$) and Na$_3$Bi ($10^{17}$--$10^{18}$)~ \cite{Feng2015,Xiong2015,He2014}. Figure. S6 (c-f) shows the evolution of carrier concentration and carrier mobilities within the temperature range 2 K – 300 K for 5 nm and 20 nm film.

Together, these magnetotransport signatures, e.g., WAL, linear MR, and carrier compensation, provide compelling evidence of spin–orbit-coupled, phase-coherent transport in Bi/EuO, marking an essential step toward experimentally establishing and tuning the topological properties of bismuth under magnetic proximity.

\section{conclusion}
\noindent In summary, we demonstrate the controlled growth of ultrathin bismuth on the ferromagnetic insulator EuO(111), establishing as a viable and robust experimental platform for realizing magnetic higher-order topology. Atomically resolved imaging confirms the stabilization of $\alpha$-phase bismuthene in the Bi(012) orientation with high structural quality and uniform bilayer morphology. Tunneling spectroscopy reveals a large and spatially homogeneous bulk gap of $\sim$400 meV persisting up to room temperature, consistent with a quantum spin Hall phase. Importantly, position-dependent STS measurement show progressive gap closing towards island edges, providing evidence of boundary-localized 1D helical channels. Complementary magnetotransport measurements further elucidate the evolution from surface-dominated to bulk-like conduction with increasing bismuth layer thickness, as reflected in the crossover from linear to parabolic magnetoresistance, multicarrier transport, and thickness-dependent Hall sign reversal. The coexistence of a sizable topological gap, well-defined edge states, and proximity to a ferromagnetic insulator with in-plane anisotropy fulfills the key ingredients anticipated for realizing magnetic higher-order topological phases. These results opens a clear pathway for low-temperature spectroscopic detection of topological corner states and provides a promising route to realise and control magnetism-driven topological phase transitions from first-order to higher-order topology. 

\section{acknowledgments}
S.N. acknowledges fellowship support from the Council of Scientific and Industrial Research (CSIR) (File Number: 09/086(1487)/2021-EMR-I), Govt. of India. This work is supported by DST funded project ‘CONCEPT’ under nanomission program (DST/NM/QM-10/2019). We thank Christoph Tegenkamp (TUC) for insightful discussions. S.N.  and S.M. acknowledge partial support of the Indo-German project SPARC-GIANT, Ministry of Education,
Govt. of India. The authors acknowledge the Department of Physics and the Central Research Facility (CRF), IIT Delhi, for various sample characterization facilities.

\nocite{*}

\bibliography{ref}

\end{document}